\documentclass[twocolumn,showpacs,amsmath,amssymb,prb]{revtex4}

\usepackage{graphicx}
\usepackage{dcolumn}
\usepackage{bm}

\begin{document}

\title{Influence of superconducting gap structure on the quasiparticle
spectrum in the vortex state}

\author{Masafumi Udagawa, Youichi Yanase, and Masao Ogata}

\affiliation{
Department of Physics, University of Tokyo, Hongo, Tokyo 113-0033, Japan
}

\date{\today}      

\begin{abstract}
We study the vortex state of a layered superconductor with vertical
 line nodes on its Fermi surface when a magnetic field is applied in the ab-plane direction. We rotate
 the magnetic field within the plane, and analyze the change of
 low-energy excitation spectrum. Our analysis is based on the microscopic
 Bogoliubov-de Gennes equation and a convenient approximate analytical
 method developped by Pesch and Dahm. Both methods give a
 consistent result. Near the upper critical field H$_{c2}$, we observe a
 larger zero-energy density of states(ZEDOS) when the magnetic field is
 applied in the nodal direction, while much
 below H$_{c2}$, larger ZEDOS is observed under a field in the
 anti-nodal direction. We give a natural interpretation to this crossover behavior
 in terms of momentum distribution of low-energy quasiparticles. We examine the recent
 field angle variation experiments of thermal conductivity and specific
 heat. Comparison with our results suggest that special care
 should be taken to derive the position of line nodes from the
 experimental data. Combining the experimental data of the specific heat
 and our analyses, we conclude that Sr$_2$RuO$_4$ has vertical line
 nodes in the direction of the a-axis and the b-axis.
\vspace{2.5cm}
\end{abstract}
\maketitle

\section{\label{sec:level1}Introduction}
Unconventional superconductors are one of the most important materials
in the modern condensed matter physics, as a key to understanding strong
electron correlation effect. A number of superconductors, high-T$_c$
cuprates, heavy-fermion metals, ruthenates, and organic compounds
exhibit unconventionality in the sense that the superconducting gap
vanishes somewhere on the Fermi surface, resulting in power-law
behaviors in various thermodynamic quantities. However, while many
superconductors were found unconventional, detailed gap structures are
still unexplored except for the cuprates. One of the difficulties in
clarifying the gap structures seem to lie in the lack of experimental
probes sensitive to quasiparticle momentum distribution. 

Recently the vortex states have been attracting much interest, because 
the positions of gap nodes can be detected. When a magnetic field
is applied parallel to the superconducting plane(ab-plane), various physical quantities
depend on the angle between the magnetic field and the
superconducting gap nodes. Hence, by rotating the field within the
plane and tracing the change of physical quantities, one can obtain the
information of the gap nodes. So far, thermal
conductivity\cite{yu,aubin,ocana,izawa,tanatar,izawa2,izawa3,izawa4,izawa5} and
specific heat\cite{deguchi,aoki,park} have been measured in the vortex state for a number of
layered unconventional superconductors including Sr$_2$RuO$_4$\cite{izawa,tanatar,deguchi},
CeCoIn$_5$\cite{izawa2,aoki}, $\kappa-$(ET)$_2$Cu(NCS)$_2$\cite{izawa3},
YNi$_2$B$_2$C\cite{izawa4,park} and PrOs$_4$Sb$_{12}$\cite{izawa5}. 

However, it is found that some of these experiments show the
behaviors incompatible with the results of theoretical analyses. 
For example, microscopic calculations show an existence of
vertical line nodes in Sr$_2$RuO$_4$\cite{kuroki,nomura,miyake,graf}, while the experiments suggest
line nodes run horizontally\cite{izawa,tanatar}. Probably these
discrepancies are attributed to the lack of firm theoretical basis in
analyzing the experimental data. Actually, experimental data have been
interpreted based on a phenomenological Doppler-shift method, which has
been claimed to be quite unreliable in some cases\cite{dahm}.
So far, there have been few microscopic analyses on quasiparticle
states under a magnetic field parallel to the ab-plane. In particular, 
no microscopic analysis has been done on layered
unconventional superconductors. Therefore, it is crucial to establish a
reliable theory in these systems, and give a correct interpretation
of the experiments. 

In this paper, we present a detailed study of quasiparticle density of
states in a layered superconductor under a magnetic field. We will focus on Sr$_2$RuO$_4$,
in which positions of the line nodes are still controversial.
A cylindrical Fermi surface with vertical line nodes of
f-wave symmetry is assumed, and a magnetic field is applied parallel to the ab-plane. 
We concentrate on the two cases where a magnetic field is in the
nodal direction, and in the antinodal direction.
We investigate the low-energy quasiparticle states, and apply the
results to the interpretation of the experimental data. 

Our analysis is based on the microscopic Bogoliubov de Gennes equation
and an approximate analytical method invented by Pesch and recently developped by
Dahm.

In the next section,  calculational formulation is described.
In section \ref{secre}, we will show our results and discussion on the
experiments. Section \ref{seccon} contains conclusion.

\section{\label{sec:level1}Formulation}
\label{secform}
\subsection{\label{sec:level2}Model and Some features}
First, we introduce some general features of our model. We study a
quasi-two-dimensional layered superconductor which has a cylindrical Fermi
surface with small c-axis dispersion. Thus we assume a dispersion
relation,
\begin{eqnarray}
\epsilon_{\mathbf{p}} = \frac{\mathbf{p_{ab}}^2}{2m_{ab}} - v_c\cos k_z,
\end{eqnarray}
where $\mathbf{p_{ab}}$ is the momentum in the ab-plane and the c-axis
wave number {\it k}$_z$ varies in the interval [$-\pi, \pi$].
In this system, the Fermi velocity can be written in the following form,
\begin{eqnarray}
\mathbf{v}_F = v_F(\cos\phi\hspace{1mm}\mathbf{e}_a + \sin\phi\hspace{1mm}\mathbf{e}_b + \epsilon \sin k_z\mathbf{e}_c),
\end{eqnarray}
with $v_F = \frac{P_F}{m_{ab}}$, and $v_c = \epsilon v_F$. 
Here, the azimuthal angle $\phi$ varies between [0,$2\pi$].
The c-axis dispersion $\epsilon$ exists due to a small inter-layer
hopping.

In this paper, we consider a spin-triplet superconducting order parameter
with its {\bf d}-vector directed parallel to the c-axis 
\begin{eqnarray}
\hat\Delta(\mathbf{k}) = \begin{pmatrix}
0 & \Delta(\mathbf{k}) \\
\Delta(\mathbf{k}) & 0 
\end{pmatrix}
.
\end{eqnarray}
We fix the momentum part of the order parameter as
\begin{eqnarray}
\Delta(\mathbf{k}) = \Delta_0(\hat{k}_a + i\hat{k}_b)({\hat{k}_a}^2 - {\hat{k}_b}^2),
\end{eqnarray}
and ignore field-induced symmetry change of the order parameter. 
This is the simplest model of chiral state with four-fold symmetric
vertical line nodes. In this model, nodes exist at
$|\hat{k}_a|=|\hat{k}_b|$. However, exact positions of nodes are not
important in the following discussion. Qualitative behavior of
low-energy density of states is determined by the relation between
applied field and line nodes. 

In order to study the vortex state under a magnetic field 
parallel to the ab-plane, we assume a spatial
variation of the order parameter, $\psi(\mathbf{r})\Delta(\mathbf{k})$.
Here $\psi(\mathbf{r})$ is described by the Abrikosov vortex square
lattice with anisotropic superconducting coherence lengths, 
\begin{eqnarray}
\psi(\mathbf{r})=2^{\frac{1}{4}}\sum\limits_{n=-\infty}^{\infty}\exp\Bigl(inq\frac{z}{\xi_c} - \frac{1}{2}\bigl(\frac{r_{\perp}}{\xi_{ab}} - nq \bigr)^2  \Bigr),
\end{eqnarray}
where, $\xi_c$ and $\xi_{ab}$ are the superconducting coherence lengths in
the c-axis and the ab-plane directions, respectively, and r$_{\perp}$ denotes the
coordinate for an axis which is in the ab-plane and is perpendicular to
the magnetic field. For example, $\psi(\mathbf{r})$ is equal to zero at
the position $r_{\perp} = \frac{q}{2}\xi_{ab}$ and $z =
\frac{\pi}{q}\xi_c$, representing the center of a vortex. In eq.\ (5),
the prefactor $2^{\frac{1}{4}}$ is a normalization factor to let the cell average of
$|\psi(\mathbf{r})|^2$ equal to 1. We have adopted the
Landau gauge for the vector potential as
\begin{eqnarray}
\mathbf{A}(\mathbf{r})=Br_{\perp}\mathbf{e}_c.
\end{eqnarray}

It is apparent that $\psi(\mathbf{r})$ has a periodicity with respect to
 $\mathbf{r}$, which corresponds to the vortex unit cell.
The spatial periods $L_c$(c-axis direction) and $L_{ab}$(ab-plane
 direction) are related to q as 
\begin{eqnarray}
L_{ab} = q\xi_{ab},
\end{eqnarray}
\begin{eqnarray}
L_c = \frac{2\pi\xi_c}{q}.
\end{eqnarray}
We choose q so that the ratio $\frac{L_{ab}}{L_c}$ is equal
to $\frac{\xi_{ab}}{\xi_c}$ in the whole range of magnetic
field. Then, $\psi(\mathbf{r})$ can be rewritten as
\begin{eqnarray}
\psi(\mathbf{r})=2^{\frac{1}{4}}\sum\limits_{n=-\infty}^{\infty}\exp\Bigl(2i\pi n\frac{z}{\L_c} - \pi\bigl(\frac{r_{\perp}}{L_{ab}} - n \bigr)^2  \Bigr).
\end{eqnarray}
Since the size of a vortex unit cell is inversely proportional
to the average induction B, we have the following relation
\begin{eqnarray}
\L_j(B) = \sqrt{2\pi}\xi_j\sqrt{\frac{B_{c2}}{B}}\hspace{5mm} (j=c, ab).
\end{eqnarray}

Finally, we introduce parameters and physical quantities of interest. 
Most properties in our system are determined by a single parameter,
namely the reduced order parameter $\eta(B)$ defined by
\begin{eqnarray}
\eta(B) = \sqrt{\frac{2}{\pi}}\frac{\Delta_0(B)\L_{ab}(B)}{v_F} ,
\end{eqnarray}
where $\Delta_0(B)$ is a spatial average of the order parameter.
Note that $\eta$ monotonically decreases as increasing magnetic field H
and becomes 0 when H reaches H$_{c2}$.

We are interested in how the density of states(DOS) depend on the angle
$\alpha$ between the field and the line node when a field
rotates within the plane. It is expected that the density of
states oscillates with the period $\frac{\pi}{2}$, reflecting the
fourfold symmetry of the order parameter. Therefore, we can concentrate
on the two cases: when a field is applied in the nodal
direction($\alpha = 0$), and in the antinodal direction ($\alpha =
\pi/4$). DOS is considered to take its minimum and maximum in
 one and the other of these two cases, respectively.

\subsection{\label{sec:level2}Bogoliubov-de Gennes equation}
Using above dispersion relations and spatially inhomogeneous
superconducting order parameter, we solve the Bogoliubov-de Gennes(BdG)
equation, which is considered to be the most reliable approach\cite{degennes}.
 \begin{eqnarray}
\hat{H_0}u_{\sigma}(\mathbf{r}) + \hat{\Delta}(-i\nabla)\Bigl[\psi(\frac{\mathbf{r}+\mathbf{r'}}{2})v_{\sigma}(\mathbf{r})\Bigr]\Big|_{\mathbf{r'}\rightarrow\mathbf{r}} \nonumber\\ 
 = E u_{\sigma}(\mathbf{r}),
 \end{eqnarray}
 \begin{eqnarray}
-\hat{\Delta}^*(i\nabla)\Bigl[\psi^*(\frac{\mathbf{r}+\mathbf{r'}}{2})u_{\sigma}(\mathbf{r})\Bigr]\Big|_{\mathbf{r'}\rightarrow\mathbf{r}} - \hat{H_0^*}v_{\sigma}(\mathbf{r}) \nonumber\\
 = E v_{\sigma}(\mathbf{r}),
 \end{eqnarray}
where
\begin{eqnarray}
\hat{H_0} = \frac{1}{2m_{ab}}\Bigl(\mathbf{P}_{ab} - \frac{e}{c}\mathbf{A}\Bigr)^2.
 \end{eqnarray}
Here we assumed $\epsilon \ll 1$, and neglected the effect of the
c-axis dispersion. This prescription corresponds to neglecting a
coherence along the c-axis direction, or in the quasiclassical sense,
to taking account of only the trajectories parallel to the ab-plane. We
will discuss the details of this prescription later in the next section.
Eq. (12) and (13) can be decoupled into 2$\times$2 matrix equation
for ($u_{\uparrow},v_{\downarrow}$) and ($u_{\downarrow},v_{\uparrow}$).
Since both the pairs satisfy the same equation, we can work on
only one of them, say, ($u_{\uparrow},v_{\downarrow}$).

We numerically diagonalize Eqs.\ (11) and (12) by descretizing the
coordinate $\mathbf{r}$, and obtain sets of eigenvalues $E_K$ and eigenfunctions
($u_K(\mathbf{r})$,$v_K(\mathbf{r})$). Using the obtained eigenfunctions, we calculate
DOS $\nu_{BdG}(\epsilon)$:
\begin{eqnarray}
\nu_{BdG}(\epsilon) = \sum\limits_{E_K > 0}\int\limits d\mathbf{r}[|u_K(\mathbf{r})|^2\delta(\epsilon - E_K) \nonumber\\
 + |v_K(\mathbf{r})|^2\delta(\epsilon + E_K)].
\end{eqnarray}

Due to the translational symmetry along the magnetic field, momentum parallel to
the vortices, $p_{\parallel}$, becomes a good quantum number. Hence, for each
eigenfunction, we can define an angle between the magnetic field and
quasiparticle momentum as
\begin{eqnarray}
\theta = \arctan\Biggl(\frac{\sqrt{p_F^2 - p_{\parallel}^2}}{p_{\parallel}}\Biggr) \hspace{5mm} (0\leq\theta\leq\pi),
\end{eqnarray}
where we limit the range of $\theta$ to [0,$\pi$] due to the reflectional
symmetry about the magnetic field. Since momentum normal to the
magnetic field $p_{\perp}$ is not a conserved quantity, $p_{\perp}$ has
a  finite width $\delta p_{\perp}$ for each eigenstate. Nevertheless,
$\frac{\delta p_{\perp}}{p_{\perp}}$ is much smaller than 1 (of the
order of $\frac{1}{k_F\xi_{ab}}$), not too much below $T_c$. Therefore,
we can consider $\theta$ as a well-defined quantity in the
quasiclassical meaning. Using this $\theta$, we discuss the momentum
distribution of the quasiparticles contributing to zero-energy density
of states.

\subsection{\label{sec:level2}Approximation due to Pesch and Dahm}
Before showing our results of BdG equations, let us introduce an
approximate analytical method, invented by Pesch\cite{pesch}, and
developped by Dahm\cite{dahm, gracer}. We will compare the obtained results with
this approximation. Near H$_{c2}$, spatial variation of order parameter
is small. Hence, in the Eilenberger equations, it is allowed
to replace normal component of quasiclassical Green functions $g$ by its
spatial average over a vortex unit cell. With this averaged
quasiclassical Green function, we can calculate various
observable quantities (e.g. DOS) in the averaged form over a vortex unit
cell. According to Dahm\cite{dahm}, even much below H$_{c2}$, this
approximation gives the result quantitatively in agreement with that
from the rigorous Eilenberger equation.

Here, we summarize the main results of this method. For details,
see Ref.\onlinecite{dahm}. We assume that a spatial variation of the order parameter is
described by $\psi(\mathbf{r})$, Abrikosov vortex lattice introduced
in the section \ref{secform}A. Then, the averaged density of
states $\nu_{PD}(\epsilon)$(in the unit of DOS for the normal state $\nu_0$) can
be written in the following form,
\begin{eqnarray}
\nu_{PD}(\epsilon) = \Bigl\langle \mathrm{Re} \frac{1}{\sqrt{1 + P(\mathbf{v_F}, i\omega_n\rightarrow \epsilon + i0)}}\Bigl\rangle_F,
\end{eqnarray}
where $\langle\cdots\rangle_F$ means averaging on the Fermi surface,
and in our system, $P(\mathbf{v_F}, i\omega_n)$ is written as
\begin{eqnarray}
P(\mathbf{v_F}, i\omega_n) = \frac{4|\Delta(\mathbf{k})|^2}{\pi|\mathbf{v_{F\perp}}|^2}\Bigl[ 1 -  \frac{\sqrt{2}\omega_n}{|\mathbf{v_{F\perp}}|}e^{\frac{2{\omega_n}^2}{\pi|\mathbf{v_{F\perp}}|^2}} \nonumber\\
\times \mathrm{erfc}\bigl(\frac{\sqrt{2}\omega_n}{\sqrt{\pi}|\mathbf{v_{F\perp}}|}\bigr)\Bigr],
\end{eqnarray}
where $\mathbf{v_{F\perp}}$ is the projection of the scaled Fermi velocity
onto the plane normal to the magnetic field,
\begin{eqnarray}
\mathbf{v_{F\perp}} = v_{F}(\frac{\cos\theta}{\xi_{ab}}\mathbf{e}_{ab} + \frac{\epsilon\sin k_z}{\xi_c}\mathbf{e}_c).
\end{eqnarray}
The momentum distribution of zero-energy quasiparticles is
simply given by
\begin{eqnarray}
\nu_{PD}(0, \mathbf{v_{F}}) = \mathrm{Re}\frac{1}{\sqrt{1 + P(\mathbf{v_F}, i\omega_n\rightarrow  +i0)}}.
\end{eqnarray}

\section{\label{sec:level1}Results}
\label{secre}
\subsection{\label{sec:level2}Zero-energy density of states(ZEDOS)}
First, we will discuss the density of states right at zero-energy.
In Fig.\ 1, we show our results for the $\eta$ dependence of
ZEDOS $\frac{\nu_j(0)}{\nu_0}$ for $j = n, a$, calculated
with the BdG and the PD methods. ZEDOS under a field
in the nodal direction($\alpha = 0$) is denoted as $\nu_n(0)$ and in the
anti-nodal direction($\alpha = \frac{\pi}{4}$) as $\nu_a(0)$.  Both
$\nu_n(0)$ and $\nu_a(0)$ are normalized to the normal-state value. 
\begin{figure}
\includegraphics[width=0.5\textwidth]{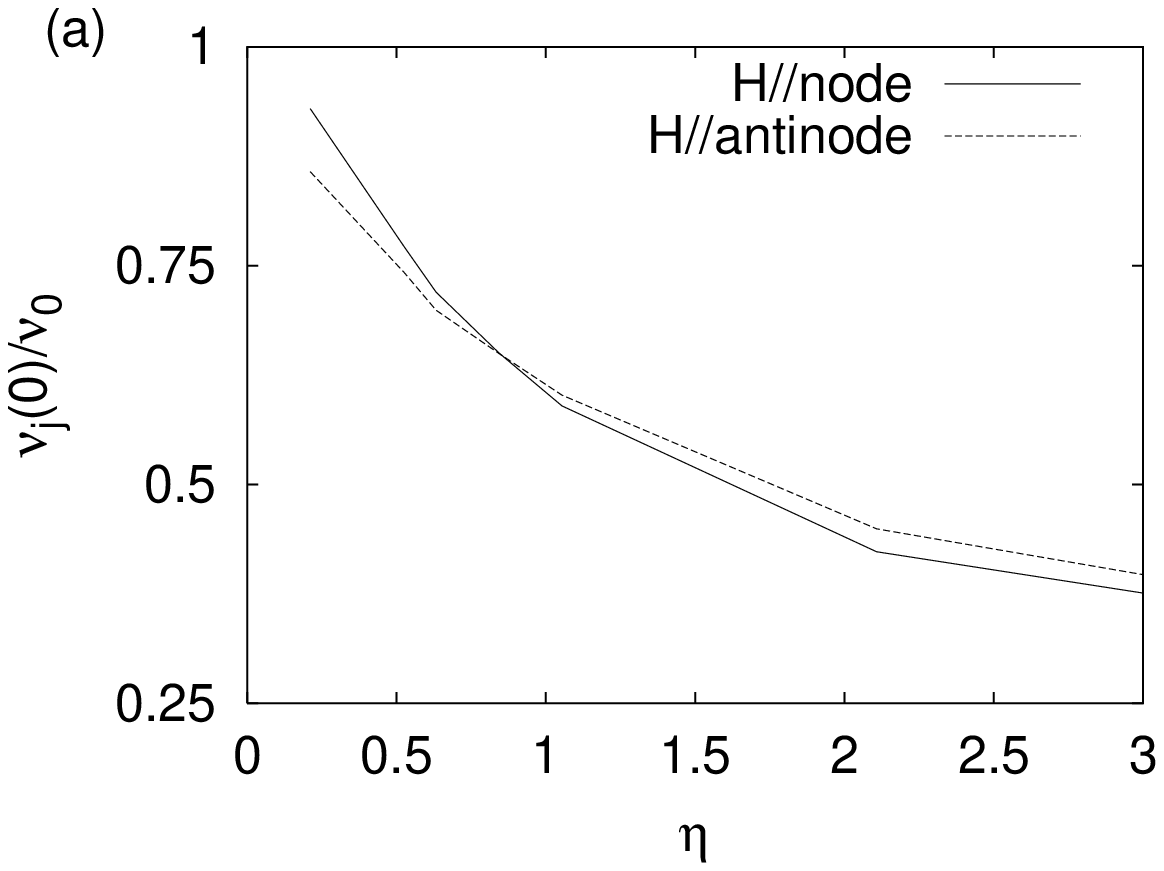}
\includegraphics[width=0.5\textwidth]{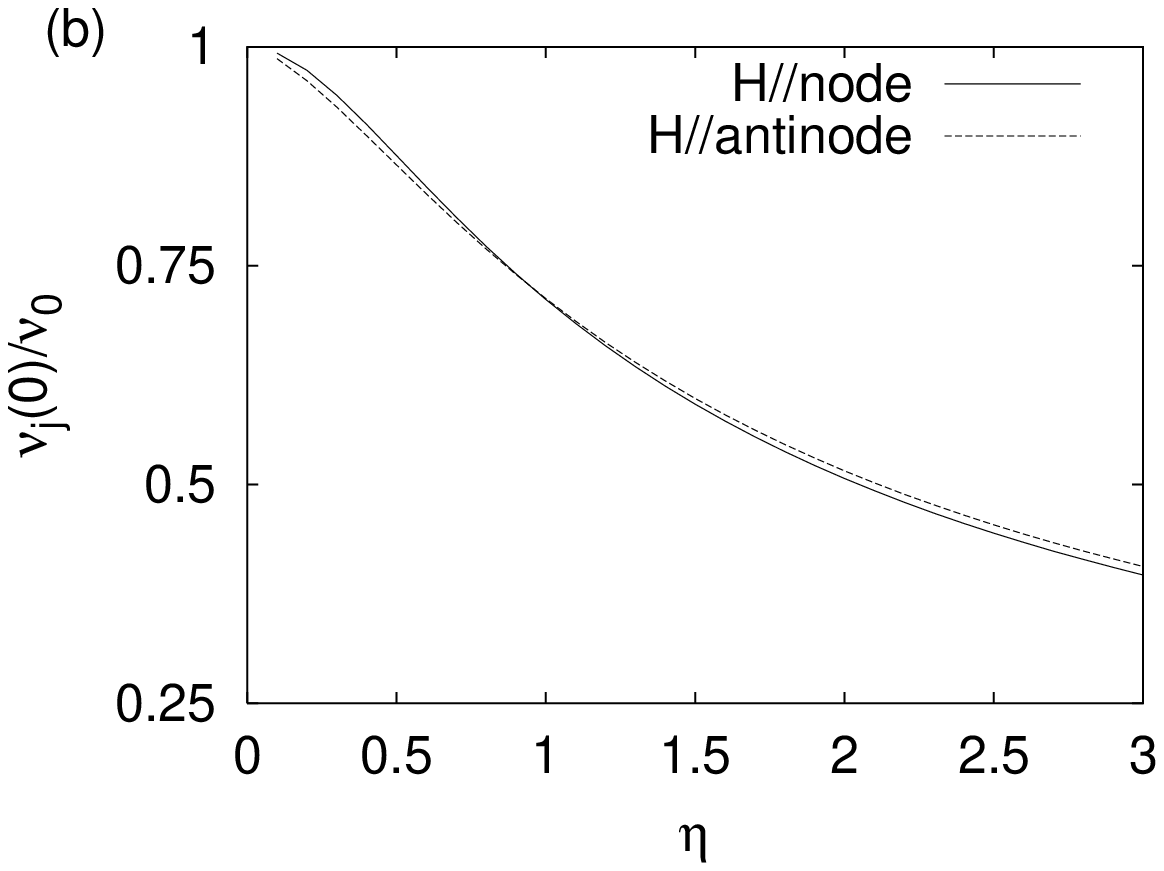}
\caption{\label{fig:}The $\eta$ dependence of zero-energy density of
 states calculated with (a)the BdG method and (b)the PD method. The solid line shows ZEDOS for
 $\mathbf{H}\parallel$ node($\alpha = 0$), i.e.$\frac{\nu_n(0)}{\nu_0}$, while the
 dashed line shows ZEDOS for $\mathbf{H}\parallel$ antinode($\alpha = \frac{\pi}{4}$),
 i.e.$\frac{\nu_a(0)}{\nu_0}$. ZEDOS is normalized to the normal-state value.}
\end{figure}

Two results show a similar behavior that in a high magnetic field
or when H is close to $H_{c2}$ (low$\eta$), $\nu_n(0) > \nu_a(0)$, while
in a low magnetic field (high $\eta$), $\nu_a(0) > \nu_n(0)$. 

In order to understand this crossover behavior, let us see the momentum
distribution of quasiparticles contributing to the zero-energy density
of states. Figure 2 shows the angle-resolved ZEDOS obtained with the BdG
method for several values of $\eta$. In a low field region (Fig. 2(a)), it is found that
narrow regions in Fermi surface are responsible for ZEDOS. In the case
of $\mathbf{H}\parallel$ node($\alpha = 0$), a sharp peak appears at
$\theta = 90^{\circ}$ which corresponds to the nodal direction
perpendicular to the magnetic field. In contrast, the
nodal quasiparticles running parallel to the field ($\theta = 0^{\circ},
180^{\circ}$) give much smaller contribution to ZEDOS. In the case
of  $\mathbf{H}\parallel$ antinode($\alpha = \frac{\pi}{4}$), we can observe nodal peaks of the same height
at $\theta = 45^{\circ}$ and $135^{\circ}$.

With increasing magnetic field (decreasing $\eta$)(Fig. 2(b)), nodal peaks become broader, and
the contribution from the core states increases by a large amount,
commonly for both $\alpha$.
This is because the decrease of order parameter amplitude makes it easier for
low-energy quasiparticles to propagate independent of $\alpha$.
However, as for the quasi-particles running parallel to
the field(i.e. $\theta = 0^{\circ}, 180^{\circ}$), we can see a big
difference between the two cases. In the case of
$\mathbf{H}\parallel$ node, contribution from the field direction becomes
larger with increasing field strength, while in the case of
$\mathbf{H}\parallel$ antinode,
contribution from this direction remains small.

When the field strength is increased ($\eta$ is decreased) furthur (Fig. 2(c)),
nodal peaks become too broad to be identified. In most part of the
Fermi surface, the angle-resolved ZEDOS recover the normal-state value. Nevertheless,
appreciable difference still exists for the quasiparticles running
parallel to the field.

We can understand the nature of this difference by using the idea of
quasi-classical trajectories. In the quasi-classical theory, a trajetory
is determined from the Fermi velocity and an impact parameter. ZEDOS can
be obtained by summing over contributions from those
trajectories. Roughly saying, if the sign of order paramter changes on a
trajectory, finite contribution to ZEDOS arises due to the formation of Andreev
bound states. Now let us consider the trajectories parallel
to the field. In the case of $\mathbf{H}\parallel$ node, quasiparticles
propagating along such trajectories feel no superconducting gap due to
the gap node irrespective of impact parameter.
On the other hand, in the case of $\mathbf{H}\parallel$ antinode, those
quasiparticles feel finite and spatially uniform gap,
because they propagate along the vortices and never cross vortex
cores. (Only part of the quasiparticles run right through a
vortex core and feel no superconducting gap. However, contributions from
such trajectories are considerably small.) Hence, in the latter case,
those quasiparticles are hampered by finite and uniform order parameter and cannot
contribute to ZEDOS, no matter how small(high) order parameter
amplitude(magnetic field) is. This is why ZEDOS is suppressed in the
case of $\mathbf{H}\parallel$ antinode in a high magnetic field region.
\begin{figure}
\includegraphics[width=0.5\textwidth]{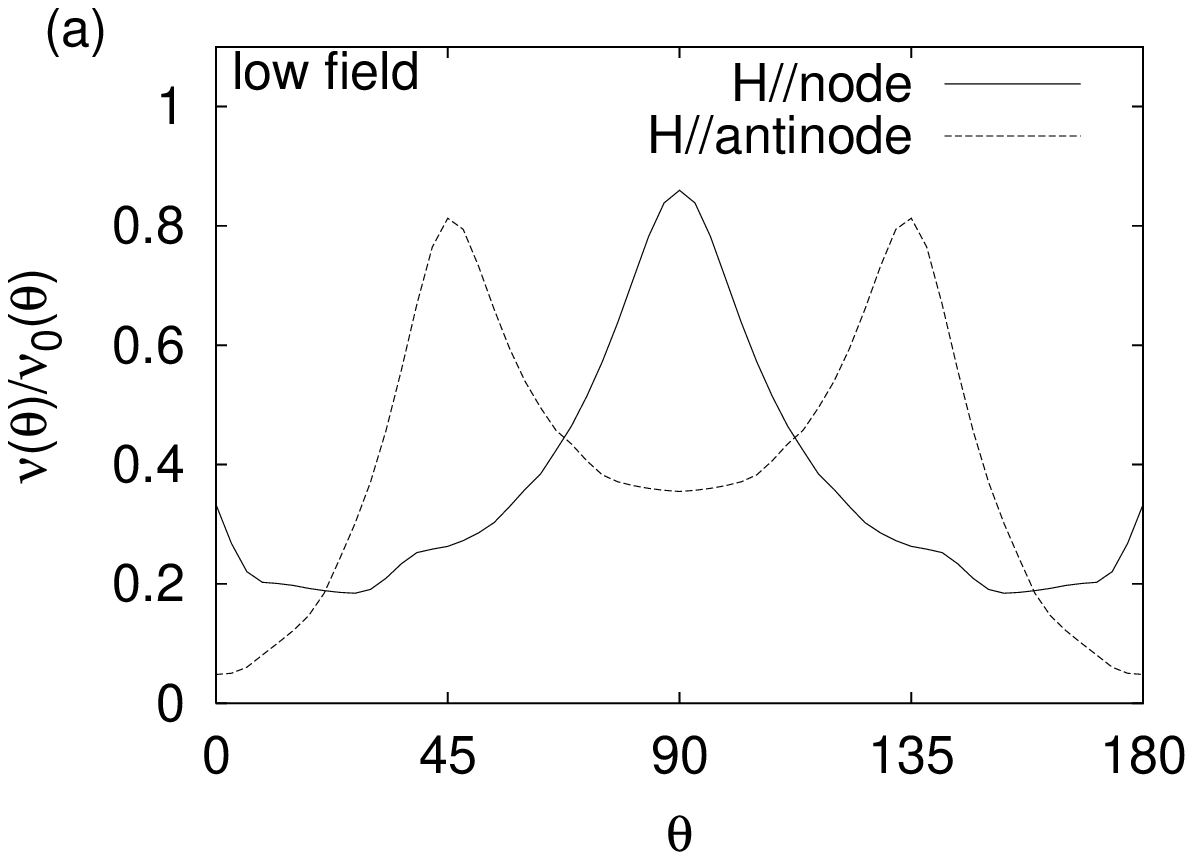}
\includegraphics[width=0.5\textwidth]{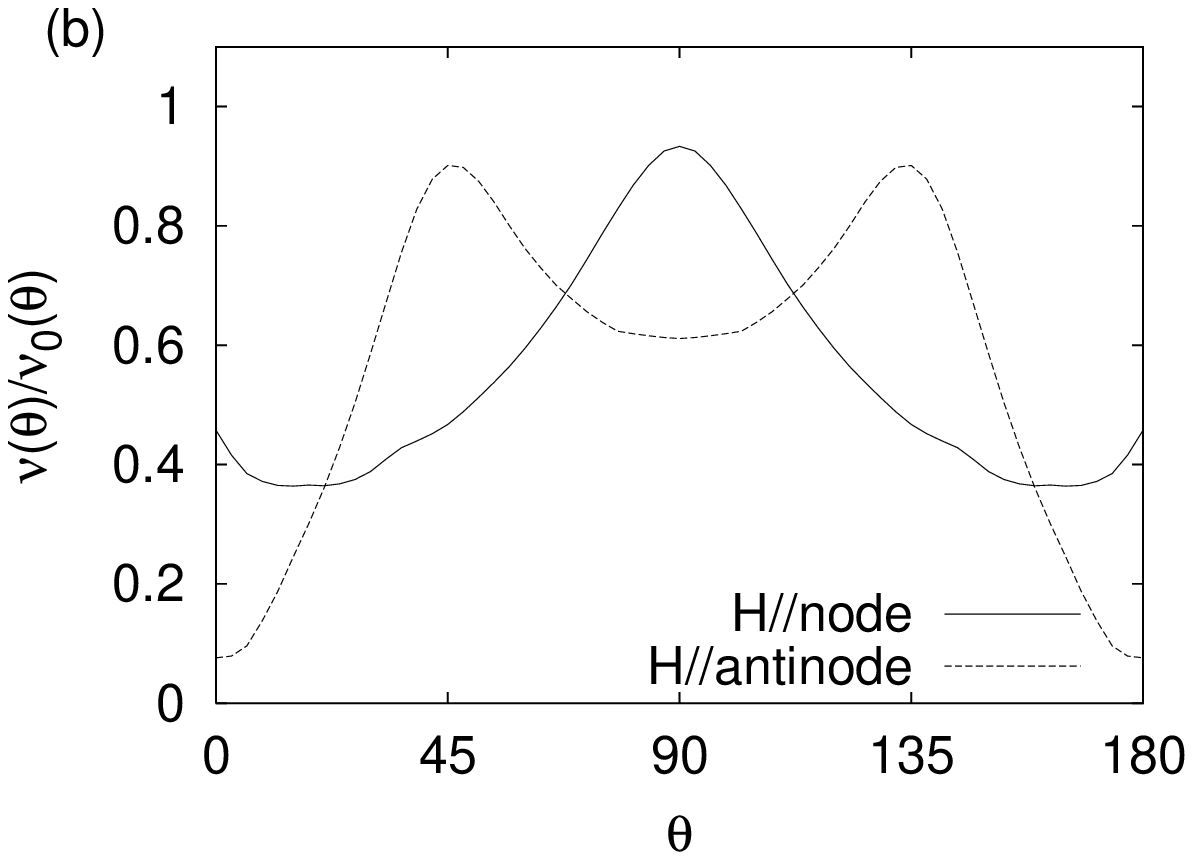}
\includegraphics[width=0.5\textwidth]{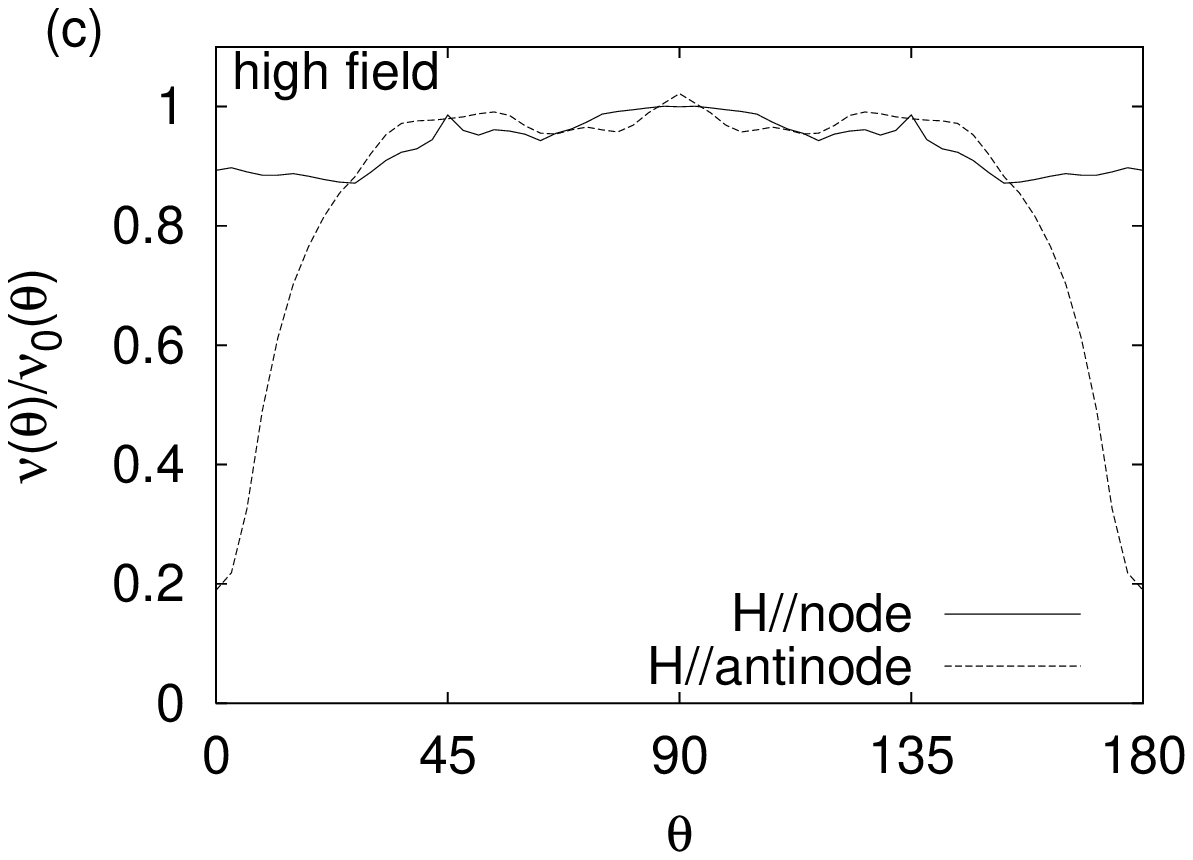}
\caption{\label{fig:zedos}Angle-resolved ZEDOS(normalized to the
 normal-state value) calculated with the BdG
 method. $\theta$ is measured from the field direction. (a)$\eta = 2.11$, (b)$\eta = 1.05$, and (c)$\eta = 0.21$}
\end{figure}

Next, let us compare the above results with the angle-resolved ZEDOS derived from the PD
method (Fig. 3) by integrating Eq.(20) with respect to $k_z$.
In most part of the Fermi surface, we observe much the same
behavior as the BdG results. However, as for the quasiparticles running
in the field direction($\theta = 0^{\circ}, 180^{\circ}$), the PD method gives larger angle-resolved ZEDOS
than that in BdG method. We attribute this difference
to the prescription we made in solving the BdG equations, that is, a neglect
of c-axis dispersion term. This prescription corresponds to considering
only the trajectories parallel to the ab-plane, i.e. limiting the c-axis
component of the Fermi velosity $k_z$ to 0. Hence, our prescription
enhances the contribution of the quasiparticles propagating
in the direction ($\theta = 0, k_z = 0$) compared with other contributions of
$k_z\not= 0$ quasiparticles. As we have discussed above, it is those
quasiparticles with $k_z = 0$ that suppress ZEDOS in the case $\alpha = \frac{\pi}{4}$.
Therefore, we obtain smaller angle-resolved ZEDOS
from the BdG method. For comparison, we plot the angle-resolved ZEDOS
obtained from Eq. (20) with $k_z = 0$ in Fig.\ 4. Comparing it
with Fig. 2(a), a quantitative agreement can be seen in $\theta\sim
0^{\circ}$ and $180^{\circ}$.

Furthur, we comment on the difference in ZEDOS observed in a high magnetic
field region($\eta\sim0.1$) between Figs. 1(a) and (b). Since the behavior of ZEDOS in
this magnetic field region is dominated by the quasiparticles parallel
to the field, our BdG analysis overestimates the difference in ZEDOS.

\begin{figure}
\includegraphics[width=0.5\textwidth]{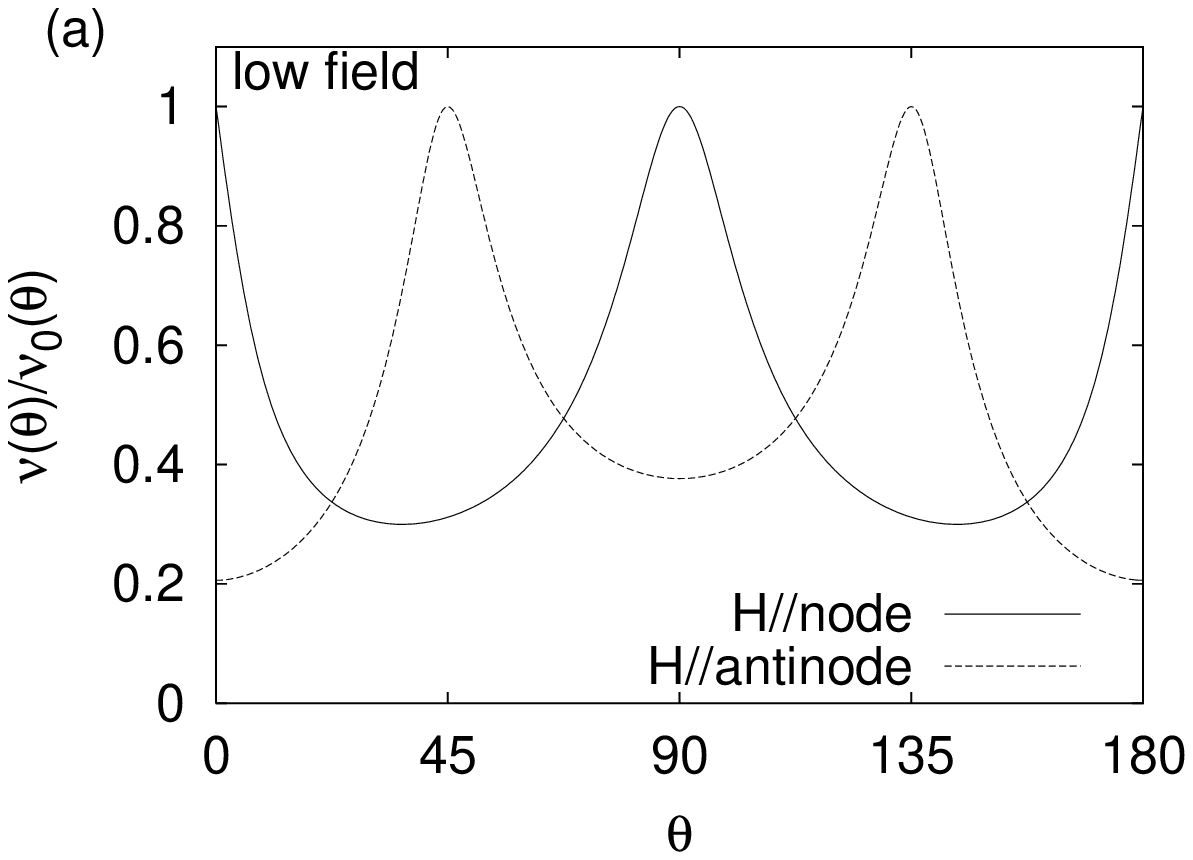}
\includegraphics[width=0.5\textwidth]{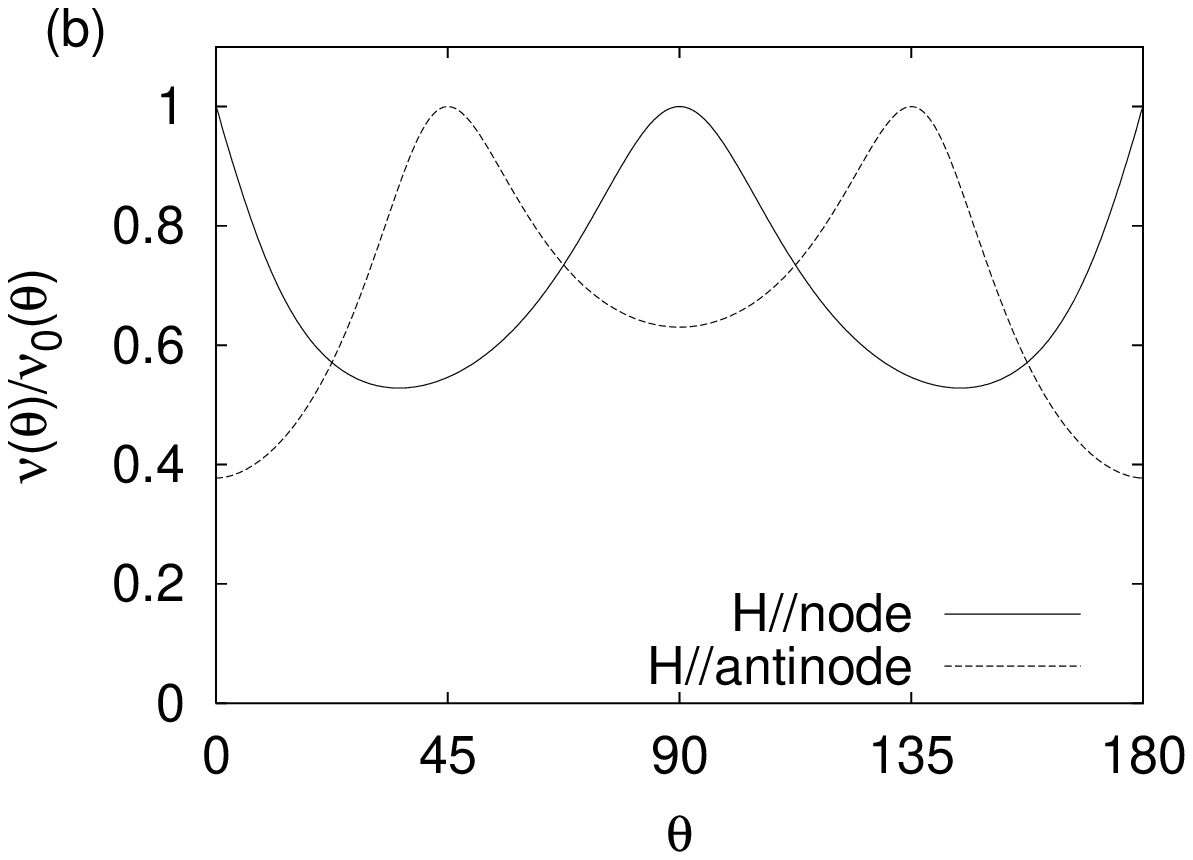}
\includegraphics[width=0.5\textwidth]{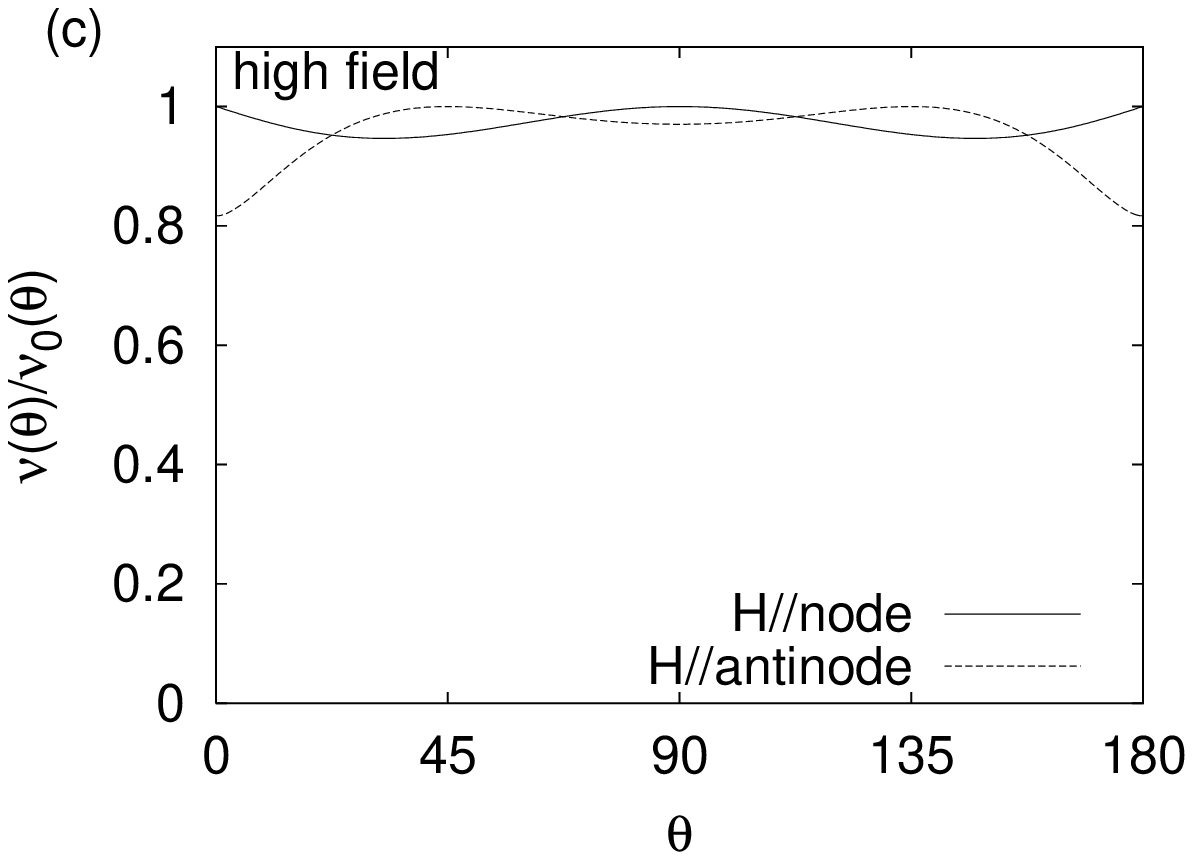}
\caption{\label{fig:zedos}Angle-resolved ZEDOS(normalized to the
 normal-state value) calculated with the PD
 method. $\theta$ is measured from the field direction. (a)$\eta = 2.11$, (b)$\eta = 1.05$, and (c)$\eta = 0.21$.}
\end{figure}
Here we briefly summarize the main results in this subsection.
In a low magnetic field region, ZEDOS is dominated by nodal quasiparticles which
have a finite momentum normal to the field. In this region,
we have $\nu_a(0) > \nu_n(0)$, since in the case of
$\mathbf{H}\parallel$ node, two of
the four nodes are parallel to the field, thus unable to contribute to
ZEDOS, while in the case of $\mathbf{H}\parallel$ antinode, all the four
nodes can contribute to ZEDOS. In a high magnetic field region, on the
other hand, the difference between $\nu_a(0)$ and
$\nu_n(0)$ comes from the behavior of quasiparticles running in the
field direction. In the case of $\mathbf{H}\parallel$ antinode, those
quasiparticles are hampered by finite and uniform order parameter and cannot
contribute to ZEDOS, while in the case of $\mathbf{H}\parallel$ node, they can.
Therefore, we have $\nu_n(0) > \nu_a(0)$ in this field region.

\begin{figure}[h]
\includegraphics[width=0.5\textwidth]{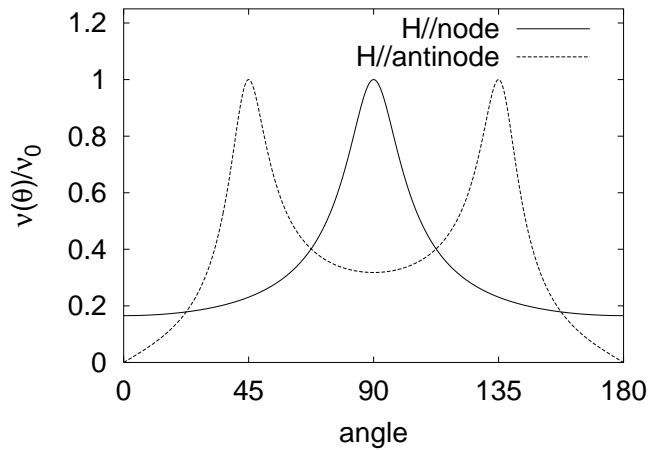}
\caption{\label{fig:zedos}Angle-resolved ZEDOS(normalized to the
 normal-state value) calculated with the PD
 method with $k_z = 0$ for $\eta = 2.11$.}
\end{figure}
\vspace{1.2cm}
\subsection{\label{sec:level2}Density of states(DOS)}
Next, we will show our results for the density of states at finite
energy. DOS at finite energy is particularly important when we discuss the
experimental data, because it depends on the density of
states at $0 \lesssim \epsilon \lesssim k_BT$.
In Figs.\ 5 and 6, we show DOS obtained with the BdG and the PD methods, respectively.

Let us first discuss the high magnetic field region (Figs. 5(b) and 6(b)), We can observe a
sharp rise in DOS at $|\epsilon|\sim\Delta_0$ reminicent of a coherence
peak in the case of $\mathbf{H}\parallel$ antinode, while not in the
case of
$\mathbf{H}\parallel$ node. This character of DOS reflects the behavior of the
quasiparticles running parallel to the field. Since they feel finite
order parameter in the case of $\mathbf{H}\parallel$ antinode, they tend
to form a coherence peak. This coherence-peak
like structure appears more clearly in the BdG result due to our
prescription as discussed in the previous subsection.

In this high magnetic field region, we observe $\nu_n(\epsilon) >
\nu_a(\epsilon)$ for $0 \lesssim |\epsilon| \lesssim\Delta_0$
independent of the calculational methods. Therefore, if the thermal
conductivity or the specific heat is measured in this field region with
a rotating magnetic field, maximum will be observed when the field is
applied in the nodal direction.

The results for a low magnetic field region are shown in Fig. 5(a) and
6(a). In this field region, as we have discussed in the previous subsection, 
$\nu_a(0) > \nu_n(0)$ at zero-energy. However, as shown in Fig. 5(a) and
6(a), $\nu_a(\epsilon)$ and $\nu_n(\epsilon)$ cross at
$|\epsilon|\sim0.2\Delta_0$. For $|\epsilon| > 0.2\Delta_0$ we have the opposite
inequality $\nu_n(\epsilon) > \nu_a(\epsilon)$. This means that when
we take experimental data at a temparature $T \gtrsim 0.2\Delta_0$
with a rotating magnetic field within the plane, we will observe very small angle
variation, because the effects of lower-energy DOS and higher-energy DOS
cancel each other.
\begin{figure}[h]
\includegraphics[width=0.5\textwidth]{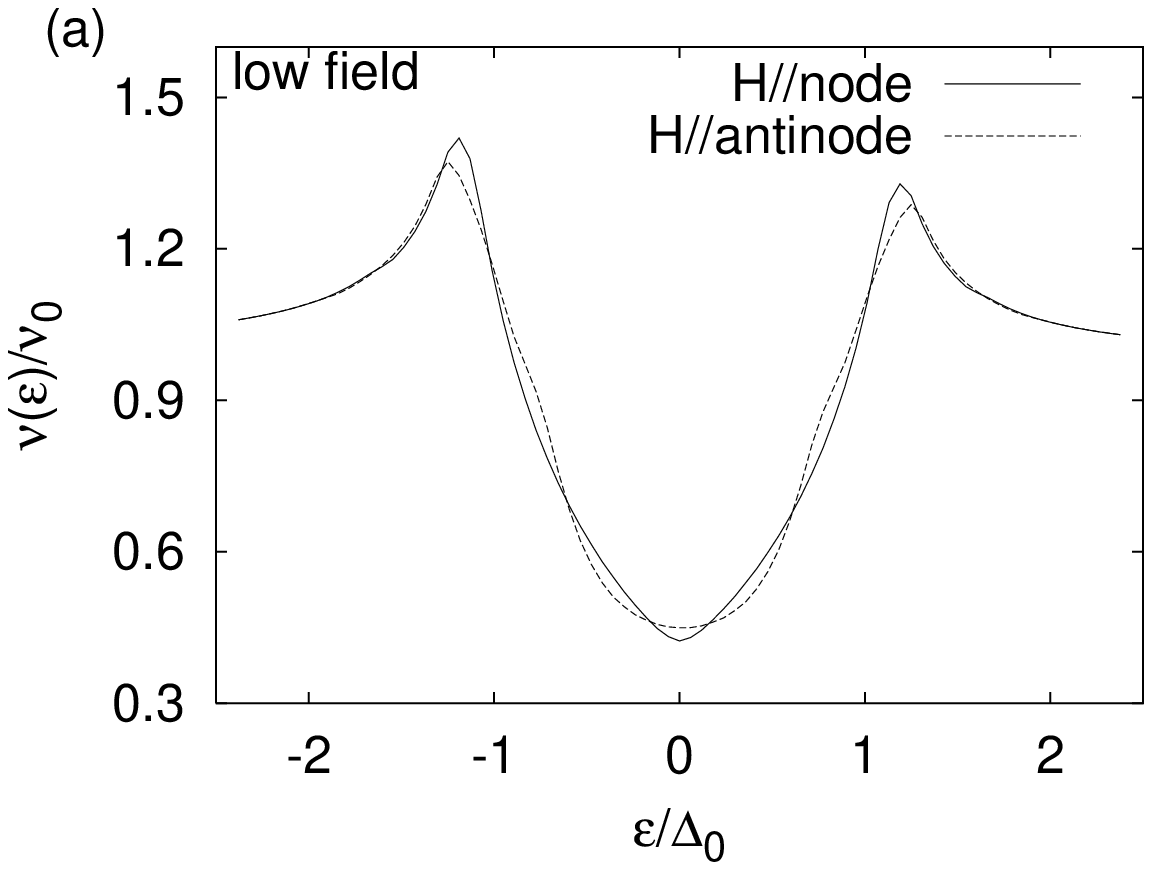}
\includegraphics[width=0.5\textwidth]{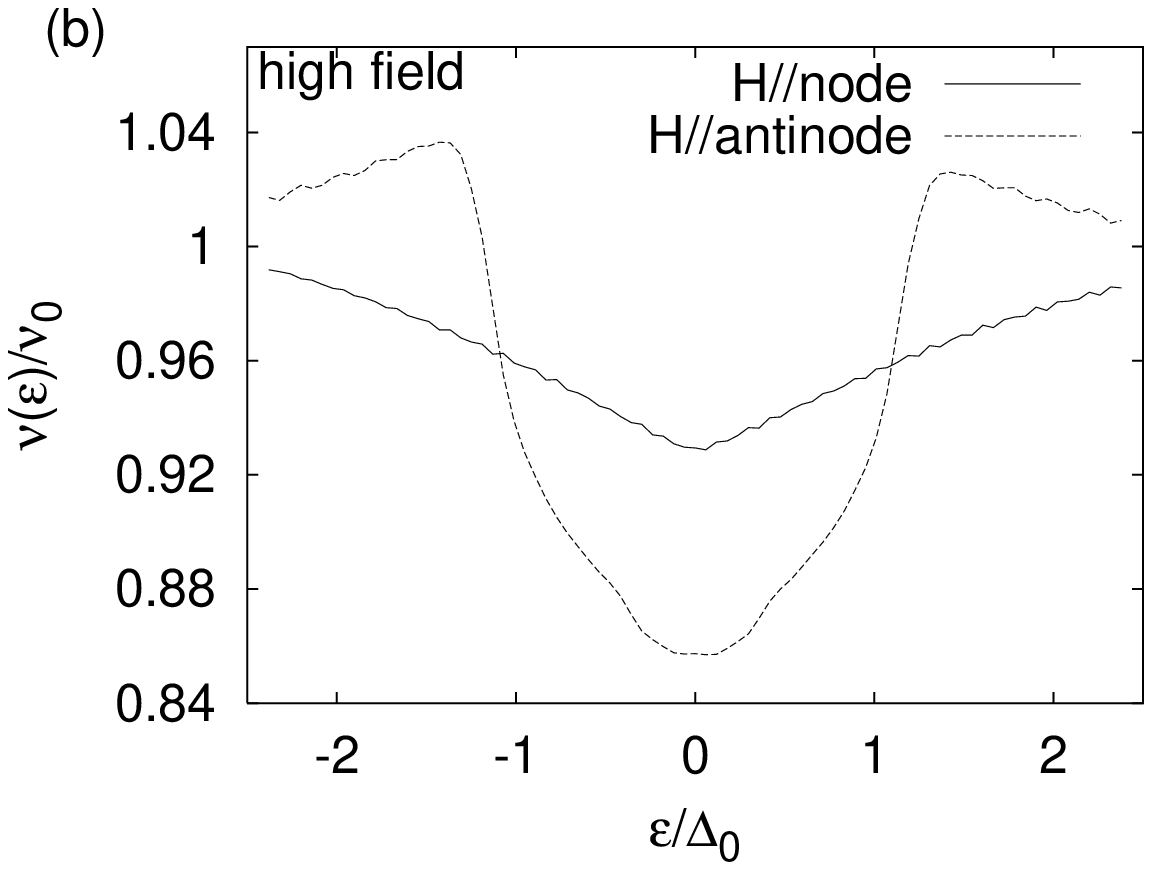}
\caption{\label{fig:zedos}Finite-energy DOS(normalized to the
 normal-state value) calculated with the BdG
 method. (a)$\eta = 2.11$ and (b)$\eta = 0.21$.}
\end{figure}
\begin{figure}[h]
\includegraphics[width=0.5\textwidth]{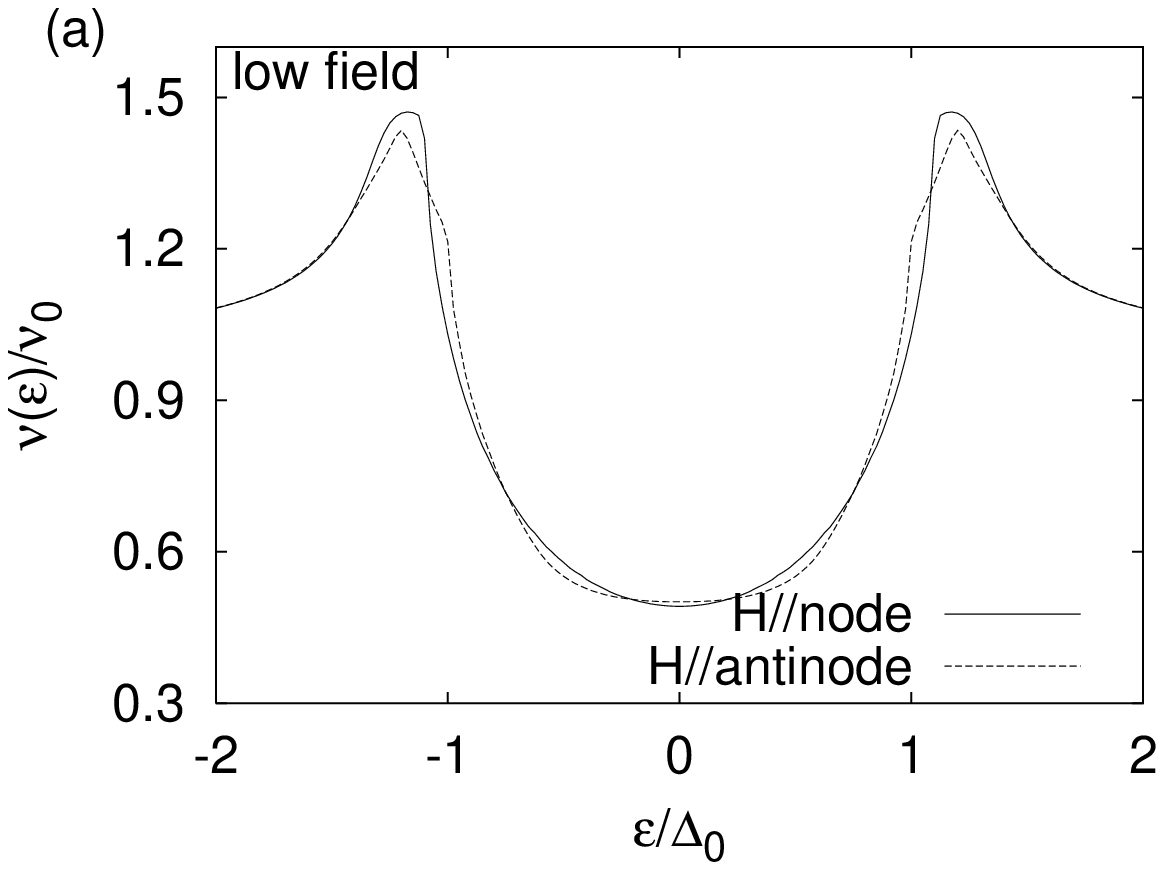}
\includegraphics[width=0.5\textwidth]{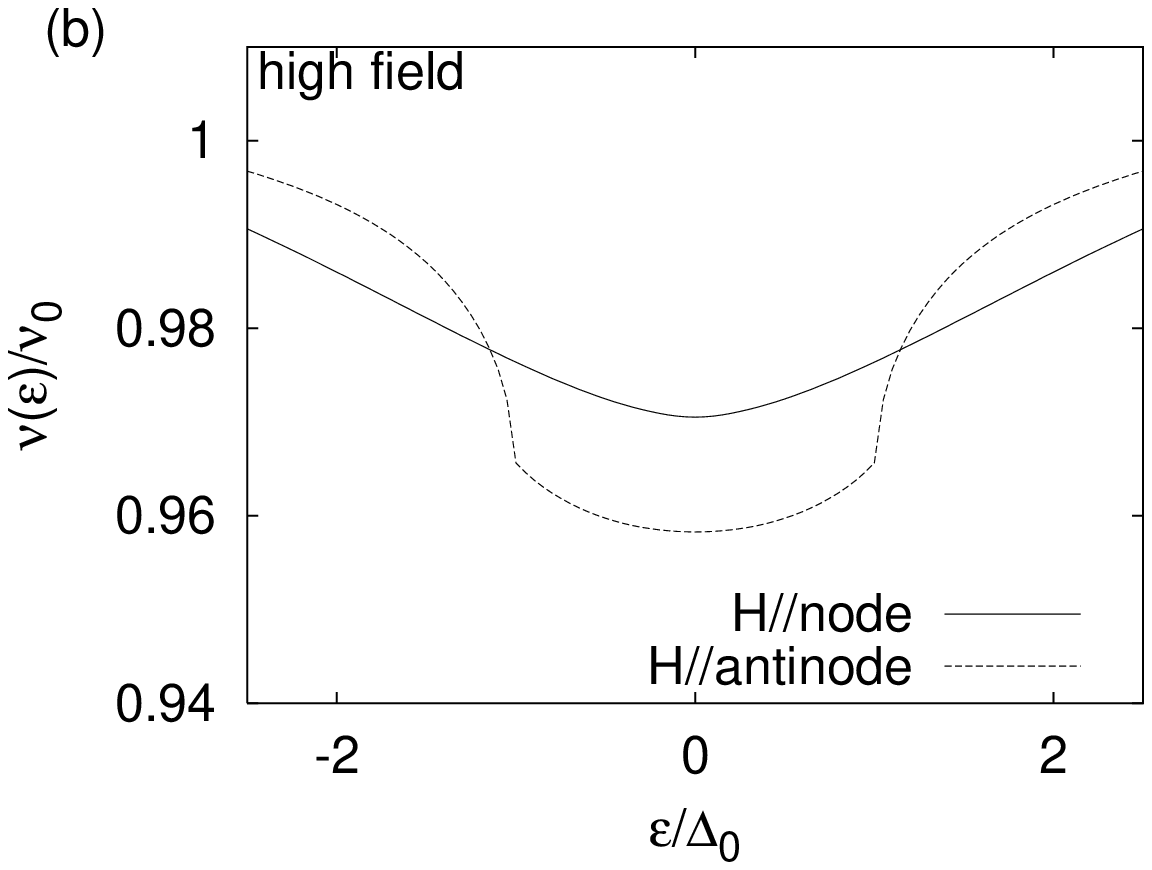}
\caption{\label{fig:zedos}Finite-energy DOS(normalized to the
 normal-state value) calculated with the PD
 method. (a)$\eta = 2.11$ and (b)$\eta = 0.21$.}
\end{figure}

 \subsection{\label{sec:level2}Interpretations of the experimental data}
In this subsection, we will discuss how to interpret the experimental data of
the thermal conductivity and the specific heat, in connection with our
analyses.

So far the thermal conductivity\cite{izawa,tanatar} and the specific
heat\cite{deguchi} of Sr$_2$RuO$_4$ have been measured under a rotating
in-plane field by several groups. In the experiments of the
magnetothermal conductivity, they found no angle variation, except in
the vicinity of H$_{c2}$. They attributed the angle variation near
H$_{c2}$ to the
anisotropy of H$_{c2}$ itself, and denied the existence of 
vertical line nodes in Sr$_2$RuO$_4$. On the other hand, in the
experiment of the specific heat, fourfold oscillation is found at a
lower field, in addition to the angle variation near H$_{c2}$. For the angle
variation at the lower field, the specific heat shows maximum at
$\mathbf{H} || $[110],
while near H$_{c2}$ maximum is observed at $\mathbf{H} || $[100].
These measurements seem to provide incompatible results. However, our
theoretical analysis gives an answer to this contradiction in the
following way.

Here we discuss why the thermal conductivity does not show the fourfold
oscillation in the low field region. As we show in the previous
subsection, the reversal of $\nu_n(\epsilon)$ and $\nu_a(\epsilon)$
occurs in this field region at $\epsilon\sim 0.2\Delta_0$. Thus, in order to observe the
anisotropy of ZEDOS in this field region, contribution from higher
energy part of DOS must be removed. However, the thermal
conductivity is measured at rather high temperatures $T = 0.2-0.3 T_c$, where the
effect of high energy part of DOS mixes inevitably as we note in the
previous subsection. Hence, it is no wonder that the anisotropy of ZEDOS cannot be observed.
On the other hand, the measurement of the specific heat was conducted at
a lower temperature $T < 0.1T_c$, where the specific heat is
sensitive to ZEDOS. Furthurmore, note that the specific heat is more sensitive to the
low energy part of DOS than the thermal conductivity.
These will be the reason why they can observe the fourfold oscillation.

Combining the above considerations, we can determine the position of line
nodes. In the measurement of the specific heat, the maximum was found at
$\mathbf{H}\parallel$[110], while from our calculations we have $\nu_a(0) >
\nu_n(0)$ in a low magnetic field region. Therefore, we conclude that
the line nodes exist in the direction [$\pm$100] and [0$\pm$10].

In sections \ref{secre}A and \ref{secre}B, we have shown that $\nu_n(\epsilon) > \nu_a(\epsilon)$
for $0\lesssim |\epsilon| \lesssim\Delta_0$ in a high magnetic field
region. Therefore, in this field region, there is a possibility that we can
observe a fourfold oscillation both in the thermal conductivity and the
specific heat, aside from the effect of the anisotropy in $H_{c2}$.

In summary, we can explain the data of specific heat and thermal conductivity
simultaneously by assuming the vertical lines nodes along [$\pm$100] axis. The line nodes
in this direction is expected in the $\gamma$-band from the relation
between the Fermi surface and crystal symmetry
(Ref.\onlinecite{miyake}), and has been confirmed in microscopic
analysis (Ref.\onlinecite{nomura}). Strictly speaking, these are not
line nodes, because excitation gap is small but finite. However, tiny
gaps serve as line nodes in the finite temperature.

\section{\label{sec:level1}Conclusions}
\label{seccon}
We studied the density of states in the vortex state of a layered
superconductor with vertical line nodes on the cylindrical Fermi
surface under a field parallel to the ab-plane. We investigated the angle
variation of DOS with changing field strength.
Bogoliubov-de Gennes equation and an approximate analytical method due to Pesch and Dahm
were solved.
We found that a field in the nodal direction gives larger zero-energy density of
states in a higher magnetic field region, whereas a field in the anti-nodal
direction results in larger zero-energy density of states in a lower magnetic field
region. This crossover phenomenon is naturally understood
in terms of the momentum distribution of quasiparticles. In a higher field
region, under a field applied in the anti-nodal direction,
quasiparticles running parallel to the field is hampered by spatially
uniform order parameter, and thus gives a smaller contribution to ZEDOS compared with
the case in which the field is applied in other directions. In a lower field region,
nodal quasiparticles not parallel to the field contribute to ZEDOS
significantly. Therefore, when the field is applied in the anti-nodal
direction, four such nodes are available, which leads to the larger ZEDOS
compared with the case of $\mathbf{H}\parallel$ node, where
only two such nodes are available. 

We investigated the angle variation of the density of states at finite energy. We found
that in a low magnetic field region, DOS at $\epsilon\gtrsim 0.2\Delta_0$ shows maximum when
the field is parallel to the node, while ZEDOS shows minimum for this
field direction. On the basis of this fine structure of DOS, we discussed
why the thermal conductivity does not show fourfold oscillation in the
low field region. Finally, combining the experimental data of the
specific heat and our analyses, we conclude that Sr$_2$RuO$_4$ has
vertical line nodes in the direction of the a-axis and the b-axis.

We are grateful to K. Deguchi for teaching us his
experimental results. we also thank N. Yoshida for helpful discussions.

\end{document}